\RequirePackage{fix-cm}
\documentclass[twocolumn,epjc3]{svjour3}
\usepackage[symbol]{footmisc}

\smartqed  
\RequirePackage{graphicx}
\RequirePackage[caption=false]{subfig}
\RequirePackage{xcolor}
\usepackage{bm}
\usepackage{xspace}
\usepackage{multirow}
\usepackage{xcolor}
\usepackage[british]{babel}
\usepackage{hyphenat}
\usepackage{amsmath,amssymb}

\usepackage{graphicx} 
\usepackage{caption}
\usepackage{subfig}
\usepackage{float} 
\graphicspath{{figures/}} 
\usepackage{booktabs}
\usepackage{hyperref}

\usepackage{eurosym}

\newcommand{\beq}{\begin{equation}}
\newcommand{\eeq}{\end{equation}}

\journalname{Eur. Phys. J. C}
\begin{document}

\title{The Mercedes water Cherenkov detector
}


\author{
        P. Assis\thanksref{LIP,IST}
        \and
        A. Bakalov\'a\thanksref{Czech}
        \and
        U.~Barres~de~Almeida\thanksref{CBPF}
        \and
        P. Brogueira\thanksref{LIP,IST}
        \and\\
        R. Concei\c{c}\~ao\thanksref{LIP,IST,e1}
        \and
        A. De Angelis\thanksref{Padova,LIP,IST}
        \and
        L. Gibilisco\thanksref{LIP,IST}
        \and
        B. S. Gonz\'alez \thanksref{LIP,IST}
        \and \\
        A. Guill\'en\thanksref{Granada}
        \and
        G. La Mura\thanksref{LIP}
        \and
        L. M. D. Mendes\thanksref{LIP}
        \and
        L. F. Mendes\thanksref{IST}
        \and \\
        M. Pimenta\thanksref{LIP,IST} 
        \and
        R. C.~Shellard\thanksref{CBPF}\stepcounter{footnote}\thanks{deceased}
        \and
        B. Tom\'e\thanksref{LIP,IST}
        \and
        J. V\'icha\thanksref{Czech}
}

\thankstext{e1}{e-mail: ruben@lip.pt}


\institute{Laboratório de Instrumentação e Física Experimental de Partículas (LIP), Lisbon, Portugal \label{LIP}
\and
 Instituto Superior T\'{e}cnico (IST), Universidade de Lisboa, Lisbon, Portugal \label{IST}
\and
Institute of Physics of the Czech Academy of Sciences, Prague, Czech Republic\label{Czech}
\and
Centro Brasileiro de Pesquisas Físicas (CBPF), Rio de Janeiro, Brazil\label{CBPF}
\and
Università di Padova, INFN and INAF, I-35131 Padova, Italy\label{Padova}
\and
Computer Architecture and Technology Department, University of Granada, Granada, Spain\label{Granada}}

\date{Received: date / Accepted: date}

\maketitle

\begin{abstract}

The concept of a small, single-layer water Cherenkov detector, with three photomultiplier tubes (PMTs), placed at its bottom in a $120^{\circ}$ star configuration (\emph{Mercedes} Water Cherenkov Detector) is presented. 
The PMTs are placed near the lateral walls of the stations with an adjustable inclination and may be installed inside or outside the water volume. To illustrate the technical viability of this concept and obtain a first-order estimation of its cost, an engineering design was elaborated.
The sensitivity of these stations to low energy Extensive Air Shower (EAS) electrons, photons and muons is discussed, both in compact and sparse array configurations. It is shown that the analysis of the intensity and time patterns of the  PMT signals, using machine learning techniques, enables the tagging of muons, achieving an excellent gamma/hadron discrimination for TeV showers.

This concept minimises the station production and maintenance costs, allowing for a highly flexible and fast installation. Mercedes Water Cherenkov Detectors (WCDs) are thus well-suited for use in high-altitude large gamma-ray observatories covering an extended energy range from the low energies, closing the gap between satellite and ground-based measurements, to very high energy regions, beyond the PeV scale.

\end{abstract}

\section{Introduction}
\label{sec:intro}

For many years  water Cherenkov  detectors have been widely used in the detection of high energy cosmic rays \cite{Haverah_Park,AugerNIMA,HAWC_tank}. Their application has been motivated by the fact that they can cover large ground surface areas at a reasonably low cost, have good timing accuracy (at the nanosecond level), and are sensitive to the charged and photon components of the Extended Air Showers. 

In this article, the concept of a small, single-layer water Cherenkov detector, with three photomultiplier tubes placed at its bottom in a $120^{\circ}$ star configuration (\emph{Mercedes} WCD) is presented. 
This detector is both sensitive to low energy EAS electrons and photons and is able to tag with good efficiency vertical or inclined muons, even in the presence of a sizeable amount of electromagnetic signal in the detector. These characteristics make it well-suited to be the critical element of an ambitious future wide-field gamma-ray observatory to be installed at high altitude (above $4\,$km a.s.l.) with an extended energy range, from many tens of GeV to many tens of PeV. Such an Observatory may thus close the gap between satellite and ground-based measurements and access a rich science program, from the observation of multi-messenger and transient events, to the probing of the extreme energy Universe and fundamental physics.

The  Mercedes WCD concept is presented in the next section, and in section \ref{sec:sim} we present the simulation framework and generated simulations sets.
In section \ref{sec:perfor} we detail the expected signals for single particles as well as the identification of single muons through the use of state-of-the-art machine learning techniques (ML), while in section \ref{sec:gh}, as an example, the expected gamma/hadron (hereafter designated as $\gamma / {\rm h}$) discrimination power of an array of WCD Mercedes detectors for TeV energies is analysed.
Finally, in section \ref{sec:conclusions}, future prospects for implementation and further developments are discussed.

\section{Mercedes WCD concept}
\label{sec:design}

The critical parameters in a WCD design are, besides the quality and transparency of the water:  the surface area; the water height; the  reflectively of the  water container walls; the type, number and the locations of the light sensors. These parameters determine in a large way the cost and the physics performance of the detector unit, namely the energy thresholds to electrons and photons, the capability to identify muons and the intrinsic time resolution on the arrival of the shower particles.

Several strategies are possible and have been successfully employed. For example, in HAWC 
\cite{HAWC_tank} the option chosen was to build WCDs with large dimensions (each HAWC WCD is a cylinder with a height of $5\,$m and a diameter of $7.3\,$ m and contains about $200\,000\,$L of water) and black walls, with the light collected by one central 10"  PMT surrounded by three 8" peripheral PMTs, all placed at the bottom of the tank. This large water height increases the difference between the mean signals produced by the muonic and the electromagnetic shower components, ensuring a high efficiency to tag muons. 
However, this strategy limits the observatory sensitivity for primary $\gamma$ rays as the large volume of water may be expensive to obtain in places where water is not readily available.

In this article, we follow the idea proposed in references \cite{ICRC_4PMT,Borja4PMTs} of having small single-layer WCDs with several PMTs at the bottom of the station to explore the signal time structures and the intensity asymmetries between PMTs of the same station. The latter are expected to be larger for muons than for electromagnetic particles. The cost per station in this type of design is driven by the cost of the light sensors. 

\begin{figure}[!t]
  \centering
  \includegraphics[width=0.3\textwidth]{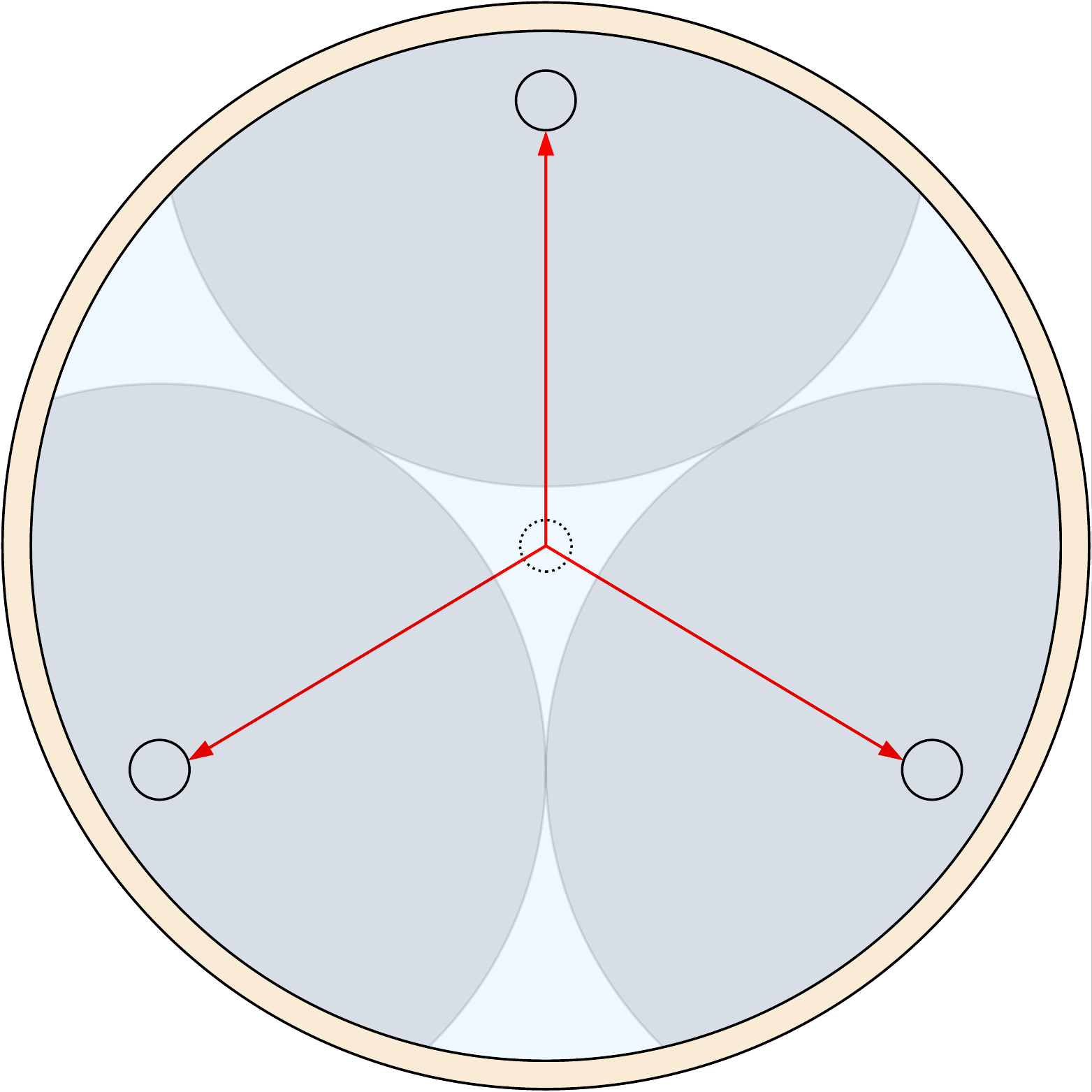}
  \caption{\label{fig:WCD-3PMTs} Bottom horizontal sketch of one Mercedes WCD. The 3-PMTs 
  are placed in a  $120^{\circ}$ star  configuration. The dark grey areas represent the Cherenkov light pools 
  of  muons injected vertically in the tank in the direction of the PMTs. 
  The dotted  small circle at  the  centre indicates the possible location of an additional  PMT placed at the top of the tank (see text). The existence of a possible thermal insulation layer connected to the lateral walls is represented in beige.
}
\end{figure}

In the Mercedes WCD, the baseline design option is a cylindrical water container with a radius of $2\,$m and a water height of $1.7\,$m and instrumented with three 8" PMTs at the bottom of the station. The dimensions of the station and the PMT positions are chosen  such that the signal asymmetry caused by a vertical muon is maximal while ensuring a complete signal coverage (see~\cite{Borja4PMTs} for details).
The PMTs are placed near the lateral walls of the stations with an adjustable inclination and in a $120^{\circ}$ star configuration (figure \ref{fig:WCD-3PMTs}), enhancing the collection of the direct Cherenkov light.
The locations of these three PMTs is such that the distance between their centres is equal to twice the radius of the Cherenkov light pool for vertical muons.

The proximity of the PMTs to the lateral walls may allow an engineering design (see section \ref{sec:conclusions}) where the PMTs would be placed out of the water volume, thus
decreasing its cost considerably and making the installation and maintenance more manageable and less costly. 
The inner walls of the container should ensure an effective light diffusion.
 In this design, the $2-3\,$ns time resolution needed to accurately reconstruct the shower geometry is achieved by the detection of the direct Cherenkov light. 
 It should be noted that the dimensions of the tank presented in this work were not optimised.

\section{Simulation framework and sets}
\label{sec:sim}

The performance at the station level was evaluated both by injecting particles directly on top of the WCD and using extensive air shower simulations. The Extensive Air Showers were simulated with CORSIKA (version 7.5600) \cite{CORSIKA} while the detector response was modeled using a simulation framework~\cite{LATTES,HAWCsim} which uses the Geant4 toolkit (with version 4.10.05.p01)~\cite{agostinelli2003geant4,Geant4_2006,Geant4_2016}.

Proton and gamma-induced shower CORSIKA simulations were used with energies respectively of $E_0 \in [0.63;6.3]\,$TeV and of $E_0 \in [1;1.6]\,$TeV and zenith angles of $\theta_0 \in [5^\circ;15^\circ]$ (vertical events). This energy range was chosen for being one of the most difficult for gamma-ray observatories wanting to rely on the tagging of muons to discriminate between gamma and hadron-induced showers.

The simulations were generated following an $E^{-1}$ spectrum as a balance between computational time and good statistics at the high energy. The experimental observation level was set at $5\,200\,$m above sea level \footnote{This altitude corresponds to the altitude of the ALMA site in Chile.}. More than $3\,000$ shower events were generated for gamma primaries, while for protons, more than $17\,000$ showers were simulated. Afterwards, a cut on the total measured WCD signal at the ground is applied to emulate a typical energy reconstruction following the procedure described in~\cite{Borja4PMTs}. In the end, one should obtain only showers which would have been reconstructed with energies around $1\,$TeV.


Both a compact and a sparse array of Mercedes WCDs were considered in this study. In the compact array, with an area of $80\,000 \, {\rm m^2}$, the WCDs are almost touching each other (fill factor of $\sim 85\%$). The sparse array is simulated by removing contiguous stations, avoiding possible shadowing effects between WCDs.

\section{Single station performance}
\label{sec:perfor}

In this section, we evaluate the performance of the \emph{Mercedes} as a detector unit either by injecting single particles at its top or using simulated shower events.

\subsection{Single particle injection}

Single photons,  with energies from $1$ to $10^5\,$MeV, and single muons, with an energy of $2\,$GeV, were injected uniformly at the top of the tank. For each simulated energy $10\,000$ photons/muons were injected. 

\begin{figure}[!t]
  \centering
  \includegraphics[width=0.5\textwidth]{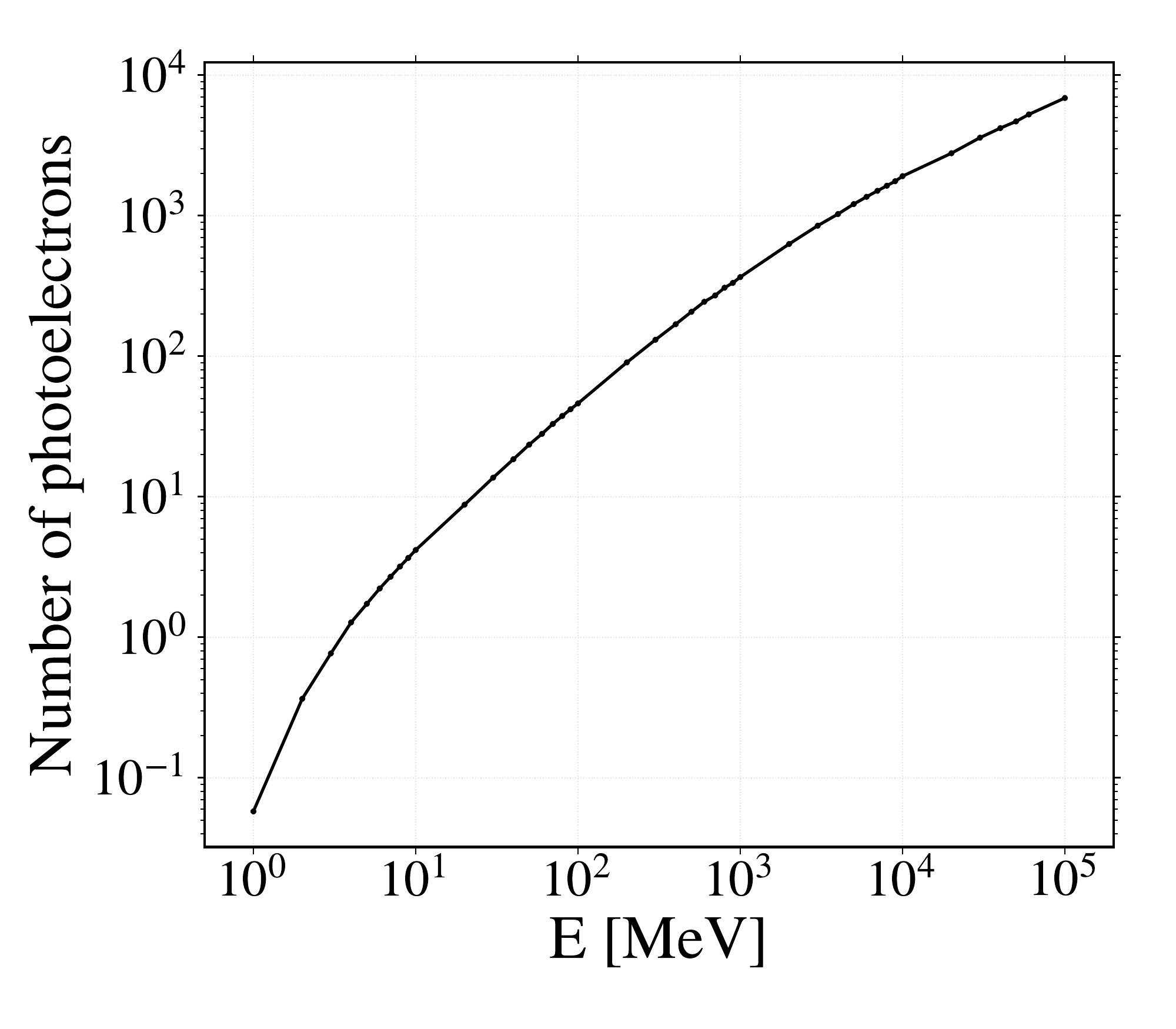}
  \caption{\label{fig:electrons}  Average gamma signal in the \emph{Mercedes} station as a function of its energy. The gamma was injected vertically and uniformly at the top of the tank. 
}
\end{figure}

\begin{figure}[!t]
  \centering
  \includegraphics[width=0.5\textwidth]{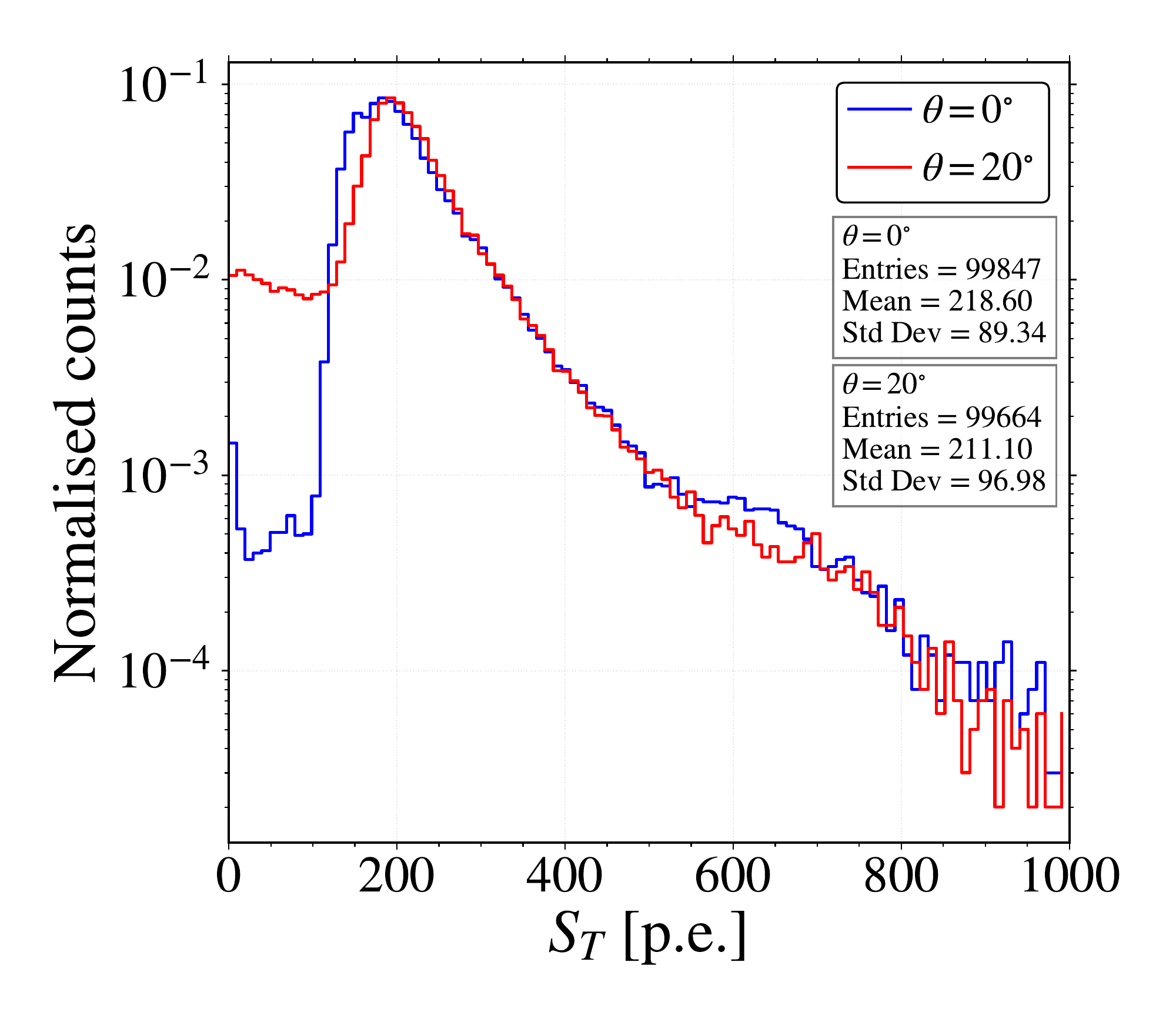}
  \caption{\label{fig:muons} Distribution of the expected total  signal of one Mercedes station for  $2\,$GeV single vertical (blue) and inclined (red) muons. 
}
\end{figure}

In figure \ref{fig:electrons} is shown the average total signal (sum of the signal of the three  PMTs) as a function of the photon energy,  while in figure \ref{fig:muons} the distributions of the total signal for both vertical (blue) and inclined (red) muons are shown.
The expected mean signal for photons of $6$, $20$ and $100\,$MeV is, respectively, about $2$, $9$ and $46\,$ photon-electrons (p.e.),  while the Vertical Equivalent Muon (VEM) signal is about $180\,$ p.e.

In figure \ref{fig:muons} the plateau in the lower region of the inclined muons is due to the clipping muons, which have a smaller path inside the tank. 
For reference, for a $1\,$TeV primary proton, the median energies of the secondary photons and muons are $\sim 6\,$MeV and $\sim 3\,$GeV, respectively~\cite{HarmJimEAS}.

\begin{figure}[!t]
  \centering
  \includegraphics[width=0.5\textwidth]{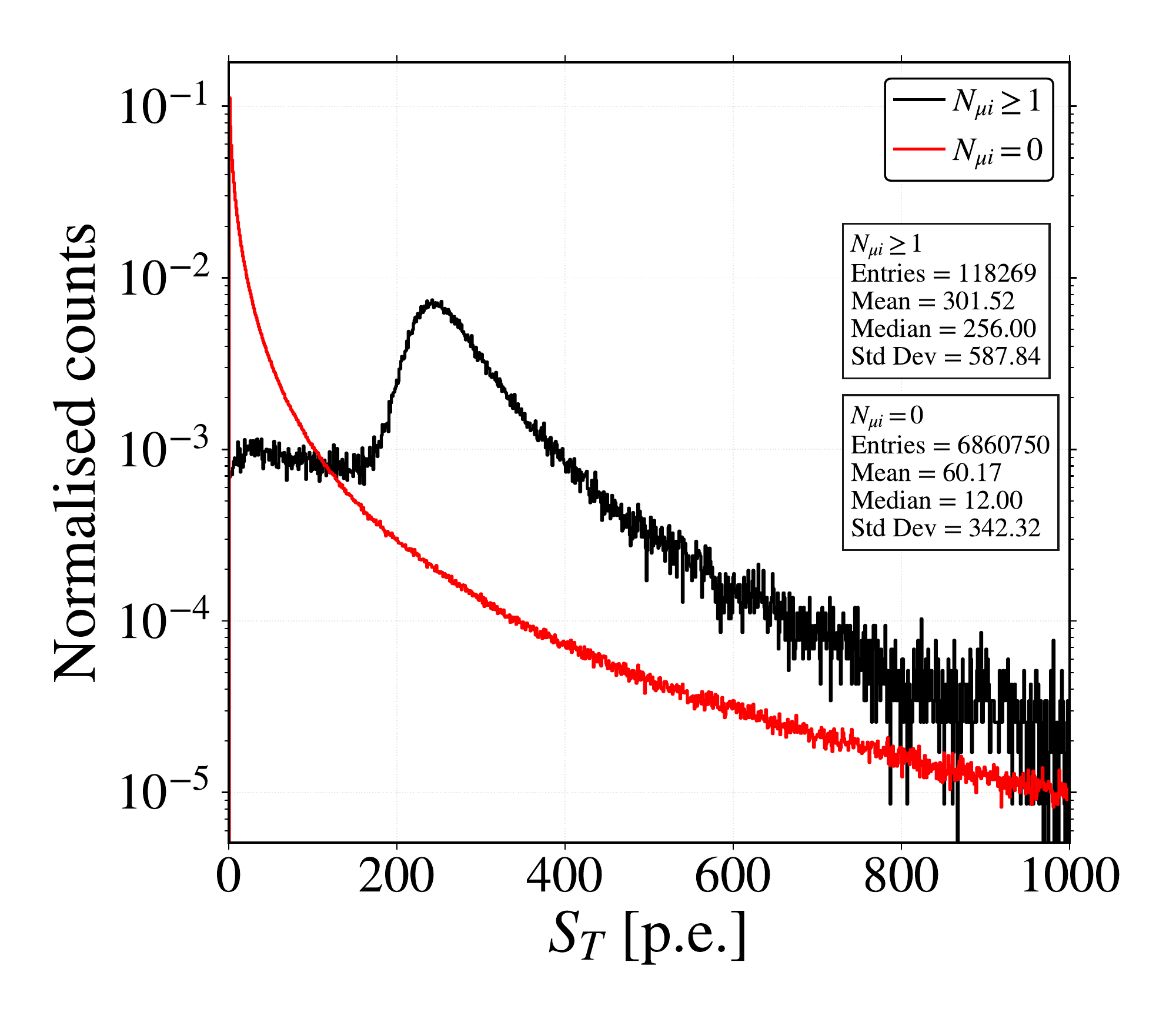}
  \caption{\label{fig:ST} Total collected signal (p.e.) in  stations with (black line) and without (red line) muons, for proton induced air showers.}
\end{figure}

\begin{figure}[!t]
  \centering
  \includegraphics[width=0.5\textwidth]{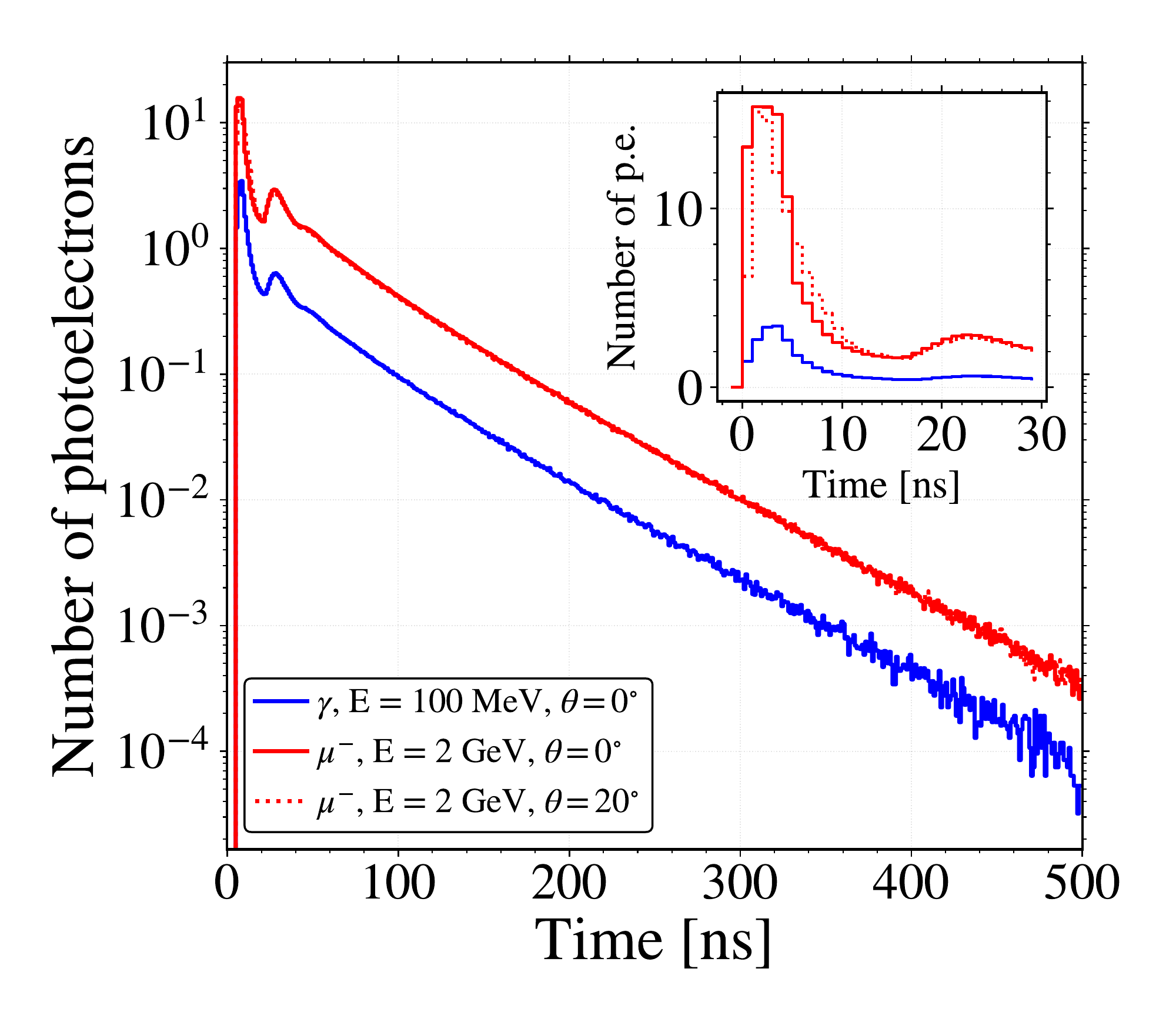}
  \caption{\label{fig:time_traces} Average PMT signal time trace for vertical 
  gammas and muons with energies of $100\,$MeV and $2\,$GeV, respectively. Inclined muons with $\theta=20^\circ$ are also shown in the plot as a dotted line. See legend for details.
}
\end{figure}

These results are in line with the requirement that the Mercedes WCD should be sensitive to low energy EAS electrons and photons while having an absolute signal for muons well above the expected electromagnetic signal in most of the stations of an array of such detectors. 
In fact, this is observed in figure \ref{fig:ST} where the distributions of the collected signals in the Mercedes stations are shown for simulations of proton-induced showers. The showers are thrown into a compact array and the stations are analysed individually to build the displayed distributions. 

Finally, in figure \ref{fig:time_traces}  are shown the average PMT signal time traces for $100\,$MeV vertical gammas and for $2\,$GeV vertical muons and inclined  muons ($\theta=20^\circ$). 
Both in the electron and the muon signal, the peak of the first arrival light is well defined (few ns), and the muon traces are identical for vertical and inclined events, enabling a good shower geometry reconstruction~\cite{LATTES}.

\subsection{Shower simulations}

\begin{figure*}[t!]
 \centering
  \subfloat[PMT$_+$]{
   \label{fig:PMT1_trace}
    \includegraphics[width=0.30\textwidth]{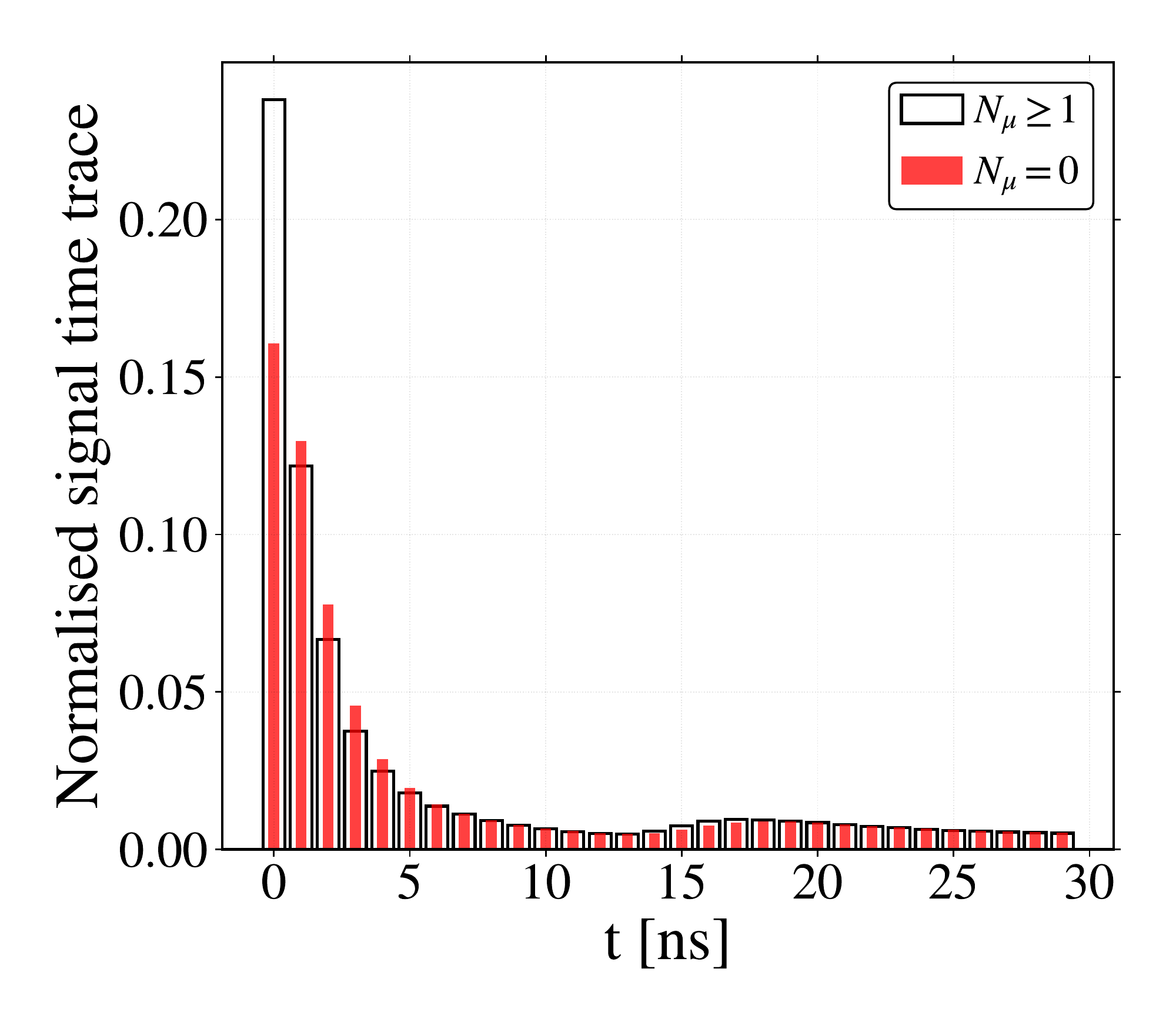}}
    \hspace{0.1in}
    \subfloat[PMT$_0$]{
 \label{fig:PMT2_trace}
    \includegraphics[width=0.30\textwidth]{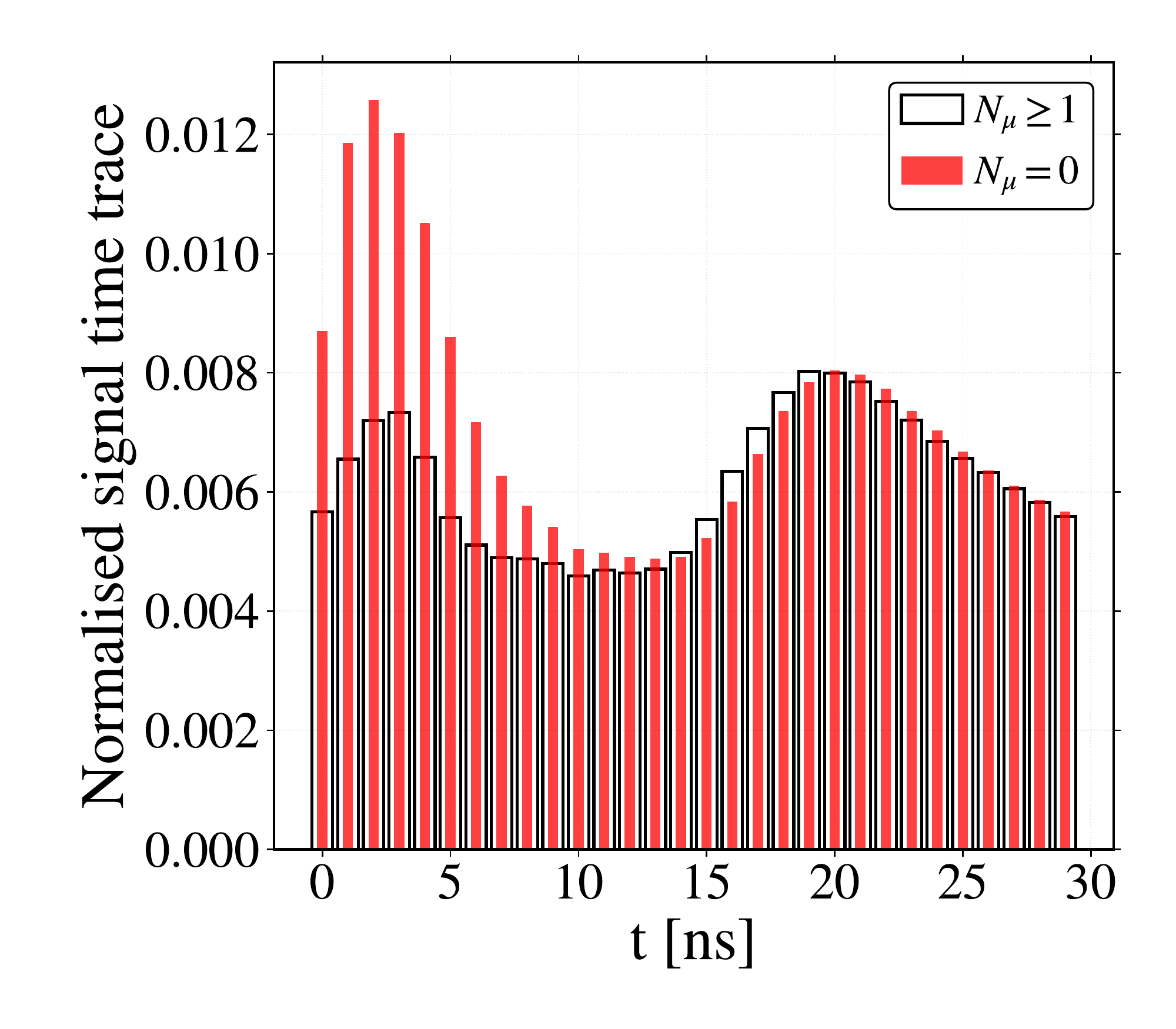}}
\hspace{0.1in}
  \subfloat[PMT$_-$]{
   \label{fig:PMT3_trace}
    \includegraphics[width=0.30\textwidth]{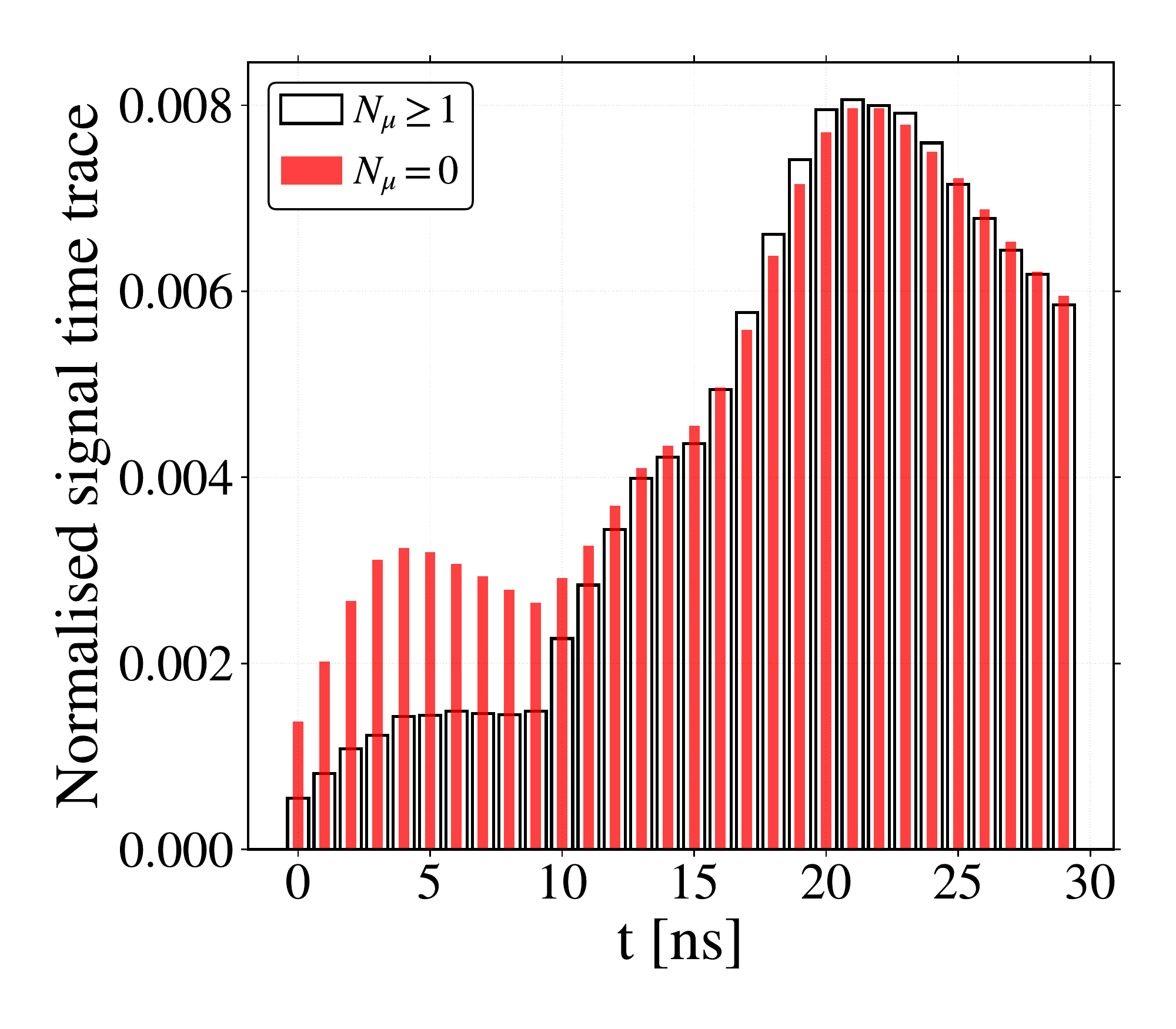}}
 \caption{Normalised mean signal time traces sorted out by their signal in the first 10 ns. In black is the trace for stations with muons and in red without muons. The PMTs were ordered and averaged according to their recorded signal, from the highest to the lowest: PMT$_+$, PMT$_0$ and PMT$_-$. 
}
 \label{fig:PMT_traces}
\end{figure*}

A large fraction of the stations with only electromagnetic particles or clipping muons may be removed by cutting signals below 200 p.e. With a similar purpose, an additional cut requires the integral up to $10\,$ns of the signal time trace of the  PMT with the highest total signal to be larger than 25 p.e.
  
The identification of muons, and of very high energy gammas, at the station level was made by analysing the spatial and temporal patterns of the PMT signals. Such is done by resorting to machine learning techniques adapting the work in~\cite{Borja4PMTs}.
  
The time traces of the signals of the three PMTs of the Mercedes WCD were analysed using a dedicated 1-dimensional Convolutional Neural Network (1D-CNN), sorting out the PMTs by their signal in
 the first $10\,$ns, the one with the highest signal being that with a higher contribution from direct Cherenkov light.

The neural network model has as inputs the following variables:

\begin{itemize}
    \item Normalised signal time traces ($0-30\,$ns) of each PMT;

    \item Integral of each PMT's signal time trace up to $10\,$ns after the start of the signal $T_0$;
    
    \item Integral of each PMT's signal time trace from  $10\,$ns to $30\,$ns after $T_0$;
    
    \item Sum of the integrals of all PMT's signal time traces  up to $10\,$ns after $T_0$;
\end{itemize}

\begin{figure}[!t]
  \centering
  \includegraphics[width=0.45\textwidth]{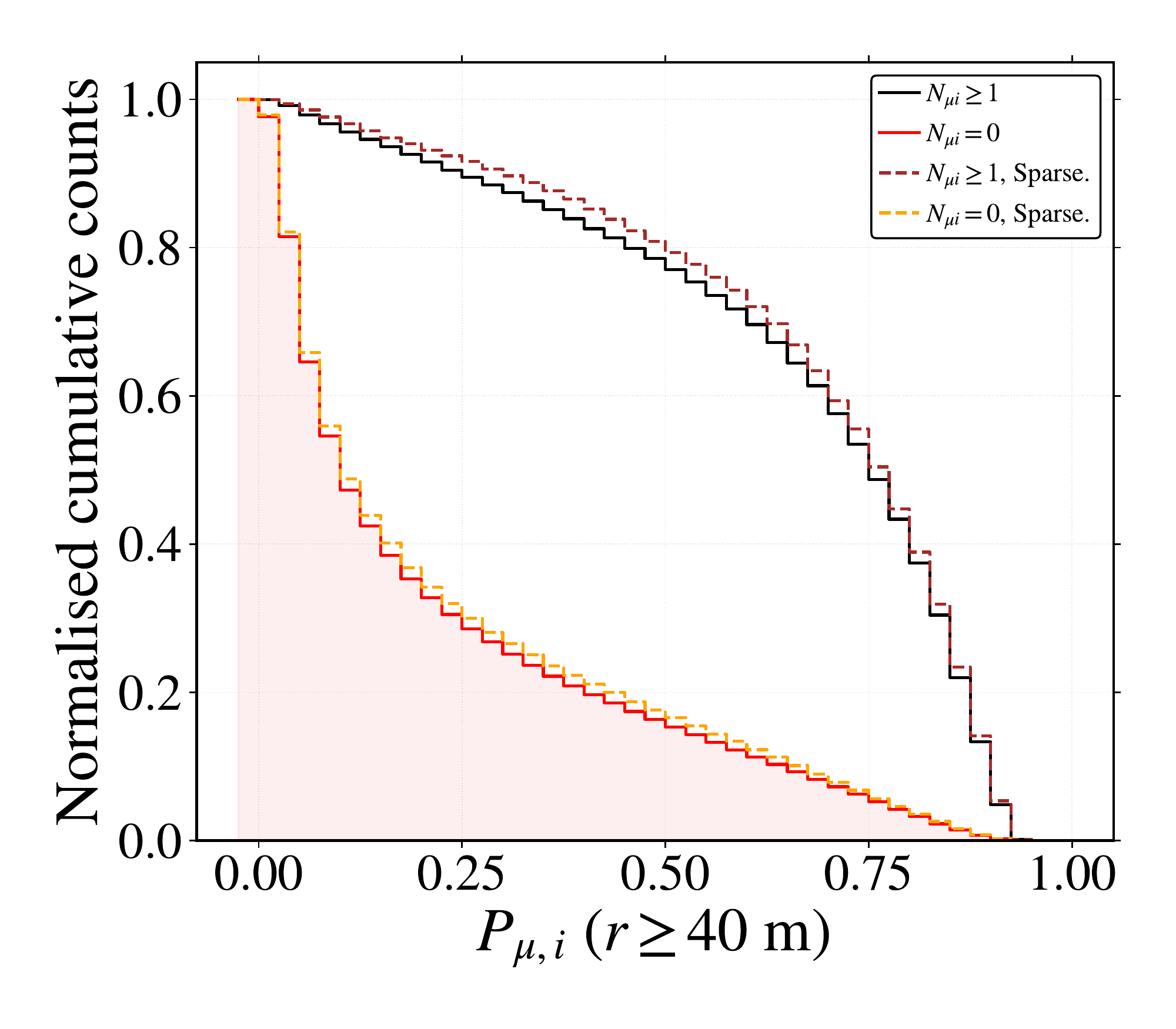}
  \caption{\label{fig:Pmui} Reverse cumulative of the distribution for the probability of finding a muon in a WCD station (the result of the CNN). Only stations at a distance higher than $40\,$m from the shower core were considered. Black line is for stations with muons while red line represents stations without muons. Results for the sparse array are also shown in the plot as a dotted line.  
}
\end{figure}

The output of this CNN is the probability $P_{\mu i}$ (figure \ref{fig:Pmui}) that the station was hit at least by one muon.

For stations at a distance higher than $40\,$m from the shower core~\cite{Borja4PMTs} and $P_{\mu i}$ greater than $0.5$, the efficiency is $80\%$, and the false-positive rate is lower than $20\%$.
The success of the CNN can be understood by building the average PMTs signal time trace ordering the PMTs in signal, shown in figure~\ref{fig:PMT_traces}. For the PMTs with more signal (a) it can be seen that stations with muons have a more prominent prompt peak than those without muons allowing the CNN to distinguish between them. Additionally, on the two remaining PMTs, it can be seen that the PMTs facing away from the entry trajectory of the shower secondary particles have an early enhanced signal for stations without muons. Such can be understood by the peaked nature of the Cherenkov light combined with a broader development of the electromagnetic shower inside the tank. Notice that the muon traverses the tank without producing any shower.

\section{Gamma/hadron discrimination}
\label{sec:gh}

An excellent gamma/hadron ($\gamma / {\rm h}$) discrimination may be achieved, following reference 
\cite{2021pgamma}, by computing  the value of the global variable, $P_{\gamma h}$, defined for each shower as:
\begin{equation}
P_{\gamma h}= \sum_{k = i}^{n_{stations}} {P_{\mu, i}}
\label{eq:Pgh}
\end{equation}
where the sum runs over all the  active stations placed at a distance greater than $40\,$m from the shower core.
$P_{\gamma h}$ is simply the sum of the probabilities of each individual selected stations being hit by a muon.

\begin{figure}[!t]
  \centering
  \includegraphics[width=0.45\textwidth]{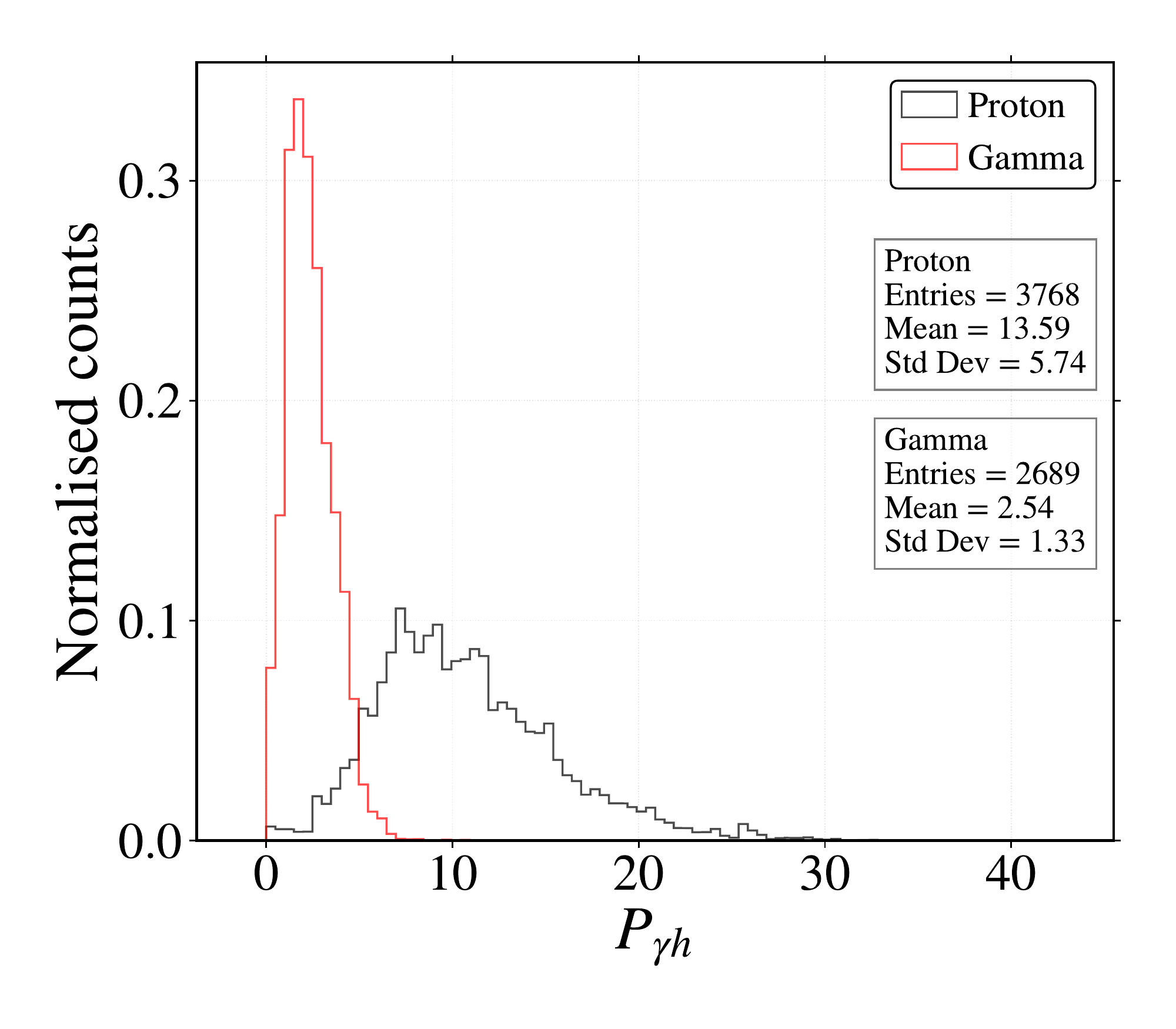}
  \caption{\label{fig:Pgh12} $P{\gamma h}$  distributions for proton (black line) and gamma (red line) events, considering the stations located at a distance greater than  40 m from the core.
}
\end{figure}

The distribution and the corresponding cumulative distribution of  $P_{\gamma h}$ for proton and gamma showers are shown in figures   \ref{fig:Pgh12}.
The values of  $P_{\gamma h}$ are, as expected, clearly higher in proton showers. The small tail at low  values for the protons are events with $E_0 \sim 0.7$ TeV, with few or even without muons at the ground.

\begin{figure}[!t]
  \centering
 \includegraphics[width=0.40\textwidth]{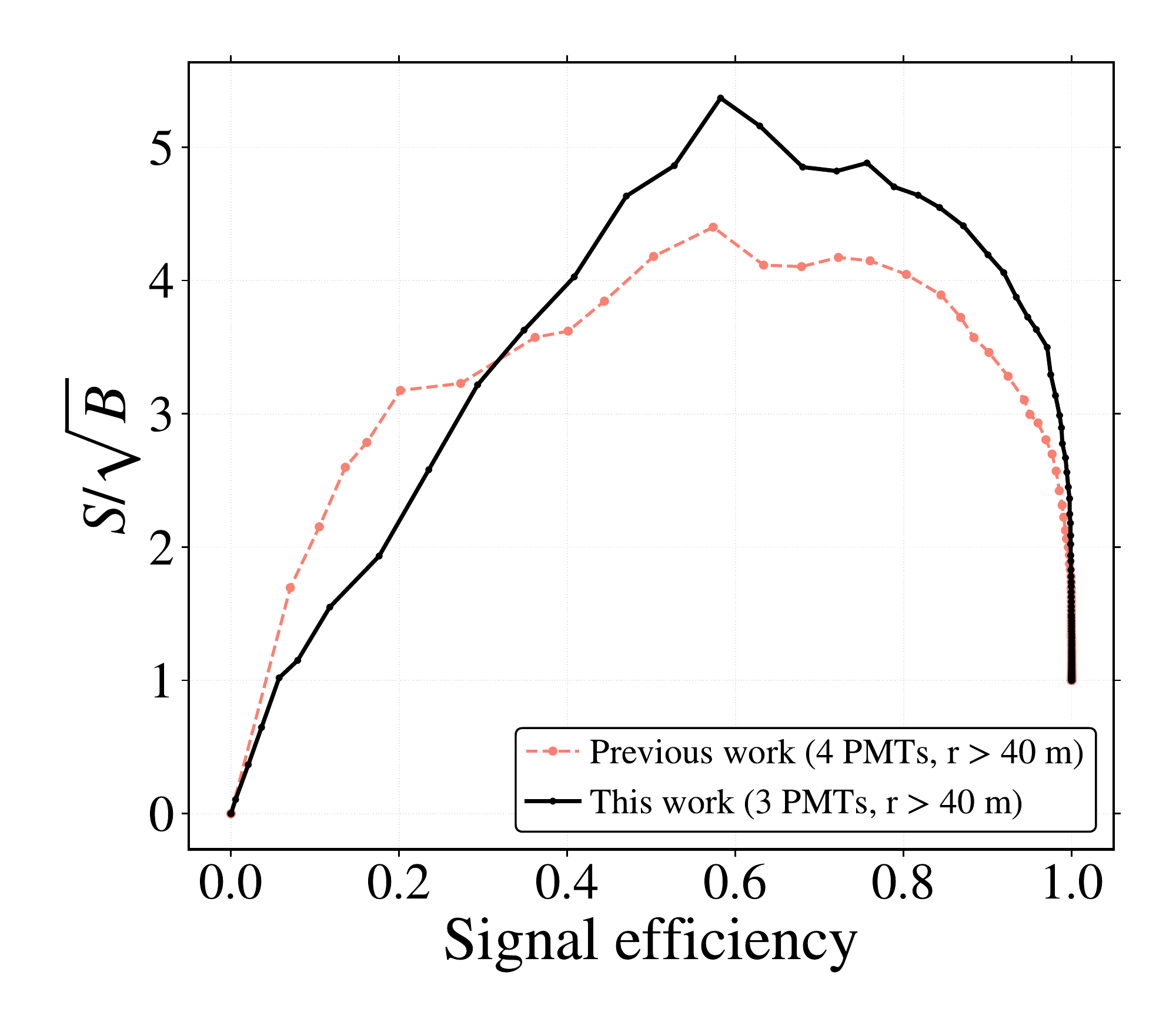}
  \caption{\label{fig:SsqrtB} Ratio of the selection efficiency for gammas ($S$) over the square root of the selection efficiency for protons ($B$) as a function of $S$. The dashed line corresponds to the result obtained with $4$ PMTs in a previous work \cite{Borja4PMTs}.
}
\end{figure}

To quantify the gamma/hadron discrimination capability, we compute the quantity $S/\sqrt{B}$, where $S$ and $B$ are the selection efficiency for gammas and protons, respectively.
The distribution of $S/\sqrt{B}$ as a function of $S$ is shown in figure~\ref{fig:SsqrtB}.
Values comfortably above 4 are observed for $S$ between $0.4$ and $0.9$, indicating an excellent gamma/hadron capability~\cite{HAWC_Crab,LATTES,Borja4PMTs}. It is important to note that while the 3-PMT curve has a higher $S/\sqrt{B}$ than the 4-PMT station, no claim should be taken from this, as it might be related to details of the optimization of the neural network. Here we aim to demonstrate that the Mercedes station and reconstruction method can be used to achieve excellent gamma/hadron discrimination; therefore, optimization studies of the algorithms are out of the scope of this work.

\section{Prospects and implementation} 
\label{sec:conclusions}
In this section, we draw some considerations about the gamma/hadron discrimination techniques and the expected performance of the \emph{Mercedes} station at PeV energies. We also discuss a possible implementation using the currently available technologies.

Contrary to what is the case at lower energies, where the discrimination variable can be readily applied as described in the previous section, at the highest energies (PeV), the number of gamma events is small, the background is huge, and the density of particles at the ground for each particle shower also significantly increases \footnote{For instance, the total number of high energy showers between $1$ and $10\,$PeV, arriving over an area of $5\,{\rm km^2}$ in one year at an altitude of $5\,000\,$m a.s.l. , is, respectively for gamma and charged cosmic rays, of the order of $10^3$ and of $10^8$; the number of muons arriving at the Earth's surface at an altitude of 5 000 m a.s.l., at distances less than 1000 m from the core is, in one proton shower with an energy of $5\,$PeV, about $70\,000$.}.
Therefore,  different strategies of muon identification should be used for stations at different distances to the shower core.
In a region of a few tens of meters around the core, the signals of the PMTs saturate, and no reliable muon identification is possible. In a region between a few tens of meters and a few hundreds of meters, where the deposited electromagnetic energy per station is higher than some hundreds of MeV, the   
identification of muons based on the signal time structures is also not efficient. However, a statistical technique based on the increase of the total signal detected in the station, due to the presence of one or more muons, as compared with the signal recorded in stations at similar distances to the core of the event \cite{2021pgamma}, may be used. Finally, a muon identification as the one described at TeV energies is again possible in the region above a few hundred meters. 
The Mercedes WCD design would be a good option in all these regions. Near the core, the addition of one small, 1-inch, low gain PMT (SPMT), placed in the centre of the station (see figure~\ref{fig:WCD-3PMTs}, would allow an extension of the station's dynamic range by a factor of about $300$ and explore charged cosmic ray physics at the PeV scale. These would have to be placed in only in a few hundred stations to avoid the saturation effects and recover information about the shower, like its core position, geometry and energy.
In the intermediate region, the fact that there are three PMTs per station increases the resolution of the total collected signal per station, increasing the sensitivity of the statistical method~\cite{IceCubeHadro}. In the outer region, where there is still a large fraction of stations and a reasonably high number of muons, the direct muon tagging as described above becomes again possible. 
Thus, an array based on Mercedes WCDs may be a good and flexible option to reach the needed background rejection factors~\cite{2021pgamma}.

The concept of the Mercedes WCD may be implemented using different technologies for the tank containers and/or the light sensors. The solution developed here, to illustrate the engineering viability of the Mercedes concept, as well as to be able to obtain a sound first-order estimation of what would be the involved costs, follows closely the design used by the Auger experiment \cite{AugerNIMA}, that has been taking data in the Argentinean Pampa for about fifteen years with very high efficiency.

The proposed tanks are rotomolded polyethylene vessels, and the water is contained in flexible liners. 
The liner is composed of three layers, the central one being one a thick carbon black loaded low-density polyethylene (LDPE) layer, opaque to photons. The other two are made of clear LDPE to avoid any black carbon migration into the water volume. This three-layer structure is bonded to a thick layer of DuPont\texttrademark\ Tyvek\textsuperscript{\textregistered} 1025-BL, which ensures an effective light diffusion~\cite{AugerNIMA}. 
The liner may have UV-transparent windows through which the PMTs may look into the water volume from below and/or from above. 
To comply with the possible mean negative temperatures over large periods of the year, at the high altitude sites, a double layer structure was foreseen with a thermal insulation layer of thickness to be optimised according to the site weather long-term parameters. 

The three bottom PMTs may be placed inside the water volume or installed in dedicated light-tight drawers with access from the outside for easier installation and maintenance. The installation of the PMTs inside the water volume is straightforward, but the price of water-proof PMTs is significantly higher (30-40\%), and the maintenance is not so easy. On the other hand, the installation out of the water volume on the tank bottom requires dedicated engineering R$\&$D, namely in installing rigid UV-transparent windows in the liner. An additional small-PMT may be installed out of the water volume in a dome mounted at the top central aperture of the tank, following the Auger design.

\begin{figure}[!t]
  \centering
  \includegraphics[width=0.4\textwidth]{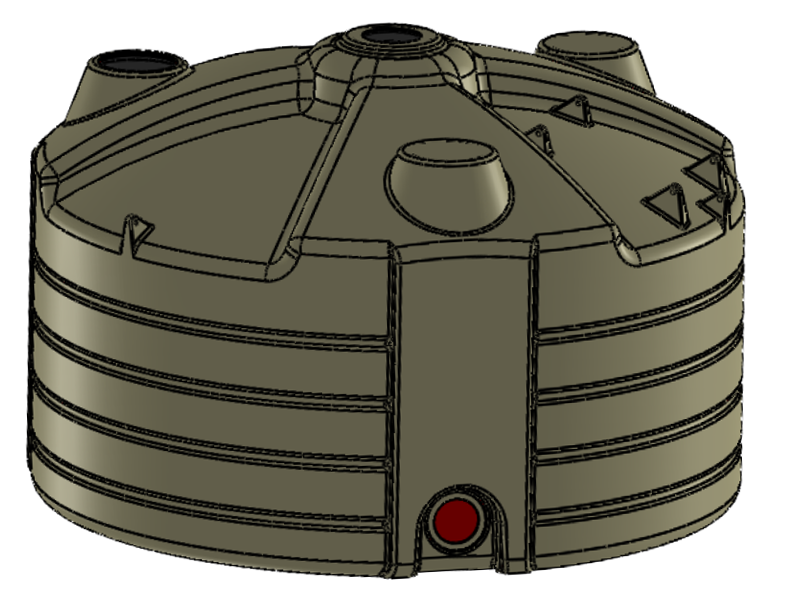}
  \caption{\label{fig:tank3D1} \emph{Mercedes} tank 3D external view.
}
\end{figure}

\begin{figure}[!t]
  \centering
  \includegraphics[width=0.4\textwidth]{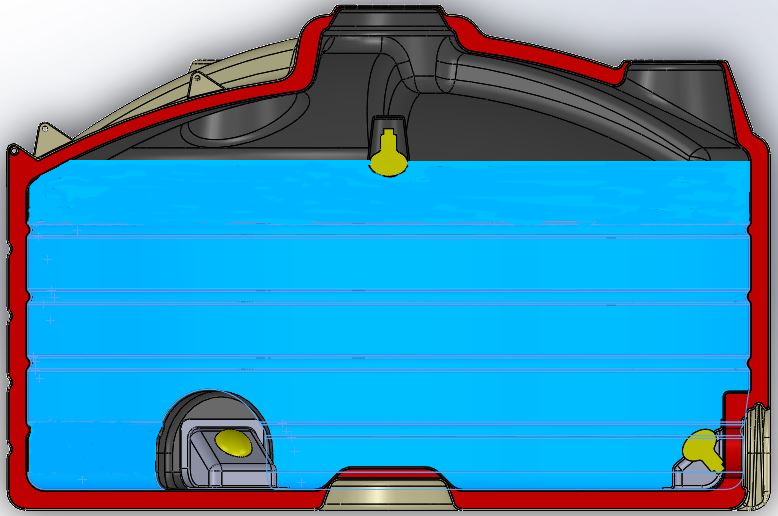}
  \caption{\label{fig:tankperfil3} View of a cut of the vertical tank profile showing its interior indicating the water level inside.
}
\end{figure}

Engineering designs, elaborated by the company Rotoplastyc, from Rio Grande do Sul, in Brazil, which has extensive experience with the production and maintenance follow-up in the field of a large fraction of the Auger tanks, are shown in figures \ref{fig:tank3D1} (3D view) and \ref{fig:tankperfil3}. The designs illustrate the implementation option where the PMTs are placed outside the water volume.

The cost of one Mercedes WCD is essentially determined by the cost of the tank, the PMTs, the liner and the water. A realistic cost estimate for this implementation based on company quotations for the rotomolded tanks and PMTs adds up to about $10.5/12\,$k\euro $\,$ (PMTs out/in the water volume) $\,$ per WCD. For the detector design where the PMT are in the water, the detector component cost can be broken down to: $3 \times 2\,$k\euro\ for the PMTs; $\sim 4\,$k\euro\ for the construction of the tank structure; $1.5\,$k\euro\ for the liner; and we took $\sim 1\,$k\euro\ for the electronics using Auger WCD electronics as a reference~\cite{AugerNIMA}. The presented numbers are the average of several quotations done at the time of writing the paper and should be taken as rough estimations.

An array based on Mercedes WCDs may thus be able to cover large surface areas at a reasonable cost with the enormous advantage of being easy to install and deploy (tanks should be ideally built at a lower altitude location near the site), and to maintain.
Furthermore,  as discussed in section \ref{sec:perfor}, the performance of the Mercedes WCD to single particles is independent of the distance to the nearest neighbour stations, allowing its efficient use both in the compact or sparse regions of the array. 
Therefore, such an array is well-suited for a flexible construction in phases according to the scientific priorities and available funding at each moment. 
In particular, such an option is presently being studied  as  one of the detector unit options for the future Southern Wide-field Gamma-ray Observatory (SWGO) \cite{SWGO} which 
will survey the Southern Hemisphere sky. 
Such an observatory will complement LHAASO~\cite{LHAASO_muon} observations, in the Northern Hemisphere, and the future Cherenkov Telescope Array~\cite{CTA}, which has a limited field-of-view (FoV) but unique pointing and sensitivity capabilities up to a few tens of TeV.

\begin{acknowledgements}
We thank our colleagues within the SWGO collaboration for many helpful discussions and the use of the common shared software framework (AERIE), which was kindly provided by HAWC. In particular, we would like to thank Andrew Smith for all the useful comments.
The authors also acknowledge the financial support by OE - Portugal, FCT, I. P., under project PTDC/FIS-PAR/4300/2020, the financial support of MEYS of the Czech Republic - grant LTT 20002 and the financial support by FAPERJ, from Rio de Janeiro State, under the Thematic Grant 211.342/2021. R.~C.\ is grateful for the financial support by OE - Portugal, FCT, I. P., under DL57/2016/cP1330/cT0002. B.S.G. is grateful for the financial support by FCT PhD grant PRT/BD/151553/2021 under the IDPASC program.  
\end{acknowledgements}

\bibliography{references.bib}   

\end{document}